\documentclass[%
superscriptaddress,
 amsmath,amssymb,
 aps,
 prb,
floatfix,
]{revtex4-2}

\usepackage{graphicx}
\usepackage{multirow}
\usepackage{dcolumn}
\usepackage{bm}
\usepackage{xcolor} %
\usepackage{mathtools}
\usepackage{hyperref}
\usepackage{tikz}
\usepackage{siunitx}
\usepackage{subcaption}

\usepackage[mathlines]{lineno}

\def\njust{Department of Applied Physics, Nanjing University of Science and Technology, Nanjing 210094, China}

\begin{document}

\title{Location of the liquid-vapor critical point in aluminum}


\author{Xuyang Long}
\affiliation{\njust}%

\author{Kai Luo}
\email{kluo@njust.edu.cn}
\affiliation{\njust}%



\date{submitted on February 9, 2026, revised on April 30, 2026}


 \begin{abstract}
The precise location of the liquid--vapor critical point (CP) in aluminum has remained elusive for decades, with reported critical 
temperatures spanning nearly 4000~K. Here we resolve this long-standing uncertainty by combining deep potential molecular dynamics 
with large-scale simulations trained on high-fidelity electronic-structure data. We benchmark multiple exchange--correlation 
functionals against experimental liquid densities and identify PBEsol as providing the most consistent description. Using 
complementary approaches---spinodal analysis of the equation of state and direct coexistence simulations with Gaussian mixture 
phase identification---we converge on a critical temperature of $6531$--$6576$~K, a critical density of $0.637$~g/cm$^{3}$, and a 
critical pressure of $1.6$~kbar. The precision of these values, with temperature uncertainties of $\sim$50~K, represents a marked 
improvement over previous estimates. Our framework establishes a transferable strategy for predicting critical phenomena in metals, 
with implications for laser ablation, shock compression, and planetary modeling under extreme conditions.
\end{abstract}

\maketitle

\section{Introduction}
 Pure metals such as aluminum, copper, gold, and platinum are of scientific interest for their unique physical and chemical properties 
 \cite{Ramana2012,Ma2013, Lemke2016,Mazhukin2022,Lentz2026}. 
 One of the most intriguing and long-standing questions about aluminum is the precise location of its liquid-vapor critical point 
 (CP). The CP marks the termination of the liquid-vapor coexistence curve, beyond which the distinction between liquid and vapor 
 phases vanishes. Determining the CP of aluminum---characterized by its critical temperature ($T_c$), critical pressure ($P_c$), and 
 critical density ($\rho_c$)---is essential both for our scientific understanding of liquid metals under extreme conditions 
 and for modeling many current high-energy-density physics experiments \cite{Vidal2001,Colombier2005,Fishburn2006, Knudson2009, Terragni2021}.
 For instance, in ultrafast 
 laser ablation simulations, spinodal decomposition occurs near $T_c$, producing distinct liquid droplets and vapor \cite
 {Vidal2001,Colombier2005,Terragni2021}. If $T_c$ is underestimated, the system can prematurely enter the supercritical regime, suppressing this two-phase 
 behavior; conversely, an overestimated $T_c$ extends the coexistence region and delays vaporization. Similarly, in shock and 
 release processes, the CP governs whether expanding aluminum crosses into the two-phase region. Under strong, multi-megabar 
 shocks, the release adiabat can approach the vapor dome, where transport properties such as conductivity and opacity change abruptly. 
 Thus,  the precise location of the CP critically influences plasma formation and dynamics, depending on whether the system 
 trajectory remains inside or passes outside the vapor dome \cite{Knudson2009}.

Despite its importance and extensive efforts, aluminum's critical temperature and pressure are poorly constrained, as summarized 
in Ref. \cite{Morel2009}. Estimates of the critical temperature ($T_c$) range
from 5115 \si{K} to 9500 \si{K}, and critical density ($\rho_c$)  spans
from 0.28 \si{g/cm^3} to 1.03 \si{g/cm^3} \cite{Young1971,Likalter1996,Hess1998,Singh2006,
Ray2006,Lomonosov2007,Povarnitsyn2008,Gordeev2008,Faussurier2009}. Some of them do not even provide critical pressure, $P_c$.
On the one hand, experimentally, direct measurements of aluminum's 
critical point are challenging due to the difficulty in maintaining and 
diagnosing under static conditions the high temperatures and pressures 
required. Therefore, most results rely on assumed rules for all liquid 
metals or  the extrapolation of thermodynamic data obtained at lower 
temperatures \cite{Morel2009}.
On the other hand, as the system approaches the critical point, thermal fluctuations
increasingly dominate. These fluctuations are notoriously difficult to capture, 
motivating the development of various theoretical approaches, including the 
renormalization-group theory \cite{Kadanoff1966, Wilson1971A, Wilson1971B, Wilson1975}.

Beyond reflecting the 
strong dependence of critical parameters on the choice of theoretical framework and computational methodology, this uncertainty 
underscores a fundamental gap in our understanding of aluminum's phase behavior. 
Such variability complicates both predictive 
modeling and experimental interpretation, emphasizing the need for reliable, 
physically consistent descriptions of aluminum's 
critical phenomena to enable unified theoretical and practical applications. 
As a result,  there is a pressing need for advanced computational methods that can 
provide more accurate predictions and insights into the critical 
phenomena of aluminum.

In this work, we address this challenge by leveraging the power of machine learning to 
develop interatomic potentials that can capture the complex interactions in aluminum with \textit{ab initio} accuracy.
With Behler-Parrinello-type neural network potentials, or deep potentials (DPs), we achieve a level of accuracy and efficiency   
that was previously unattainable \cite{Behler2007, Behler2011, Zhang2021}.
Using two complementary approaches---spinodal analysis of the equation of state (EOS) 
and direct coexistence simulations using temperature quench molecular dynamics,
we converge on a precise location of the critical point, with temperature uncertainties of $\sim$50 K.
For the direct coexistence simulations, we develop a novel microstructural analysis method based on Gaussian mixture model (GMM) 
fitting to identify the liquid and vapor densities. 
This method, described in the Supplementary Material \cite{SuppMaterial}, may also be useful in other studies of phase coexistence.

This paper is organized as follows. 
In Section \ref{sec:methods}, we describe the methodology of this work, 
including the details of the \textit{ab initio} calculations, the training 
and validation of the deep potential, the calibration of the exchange-correlation functionals, and two approaches for 
determining the critical point. 
In Section \ref{sec:results}, we present the main results of this work, predicting critical parameters using 
these two approaches: (a) the spinodal lines using an equation-of-state model in Section \ref{sub:spinodal}, 
and (b) the liquid-vapor coexistence curve in  Section \ref{sub:coex}. We also compare the predicted Clausius-Clapeyron relation 
with earlier works in Section \ref{sub:vapor_pressure}. 
Finally, we give a summary of this work in Section \ref{sec:summary}. 

\section{Methods}
\label{sec:methods}
The methodology of this work involves many technical aspects, including 
\textit{ab initio} calculations, training and validation of deep potentials, 
spinodal lines from an EOS model, liquid-vapor coexistence simulations,
 and critical point identification.

\subsection{\textit{Ab initio} calculations}
\label{sub:aimd}
All the ab initio simulations
are performed using the ABACUS package \cite{Chen2010}, which  
is efficient for large systems with numerical atomic basis sets. 
For all  \textit{ab initio}  calculations, we use the 3-electron optimized
 norm-conserving Vanderbilt pseudopotential (ONCV)
 \cite{Hamann2013} (v0.3) to describe the electron-ion interactions.
 The nonlocality of the pseudopotential is included via the standard nonlocal projectors \cite{Kleinman1982}.
Double-$\zeta$ plus polarization function (DZP) basis set is used for the valence electrons.
The energy cutoff for the real-space grid is set to 75 Ry for generalized gradient approximation (GGA) 
exchange-correlation functionals as shown in the 
ABACUS Pseudopot-Nao Square project \cite{APNSProject},
and increased to 100 Ry for the 
SCAN \cite{Sun2015} functional. 
In our study, we employ ab initio molecular dynamics (AIMD) simulations
on a grid of thermodynamic conditions
to generate accurate reference data for training the DP model.
This region covers both the liquid and vapor phases of aluminum such that the initial 
DP model can describe phase coexistence near the critical point.
Fortunately, configurations at low densities contain regions with much lower 
local density. Going further towards even lower density regime is computationally prohibitive 
for \textit{ab initio} calculations, and thus the lowest density considered is 0.3 \si{g/cm^3}.
Specifically, the low-density conditions include 
temperatures of  [3000, 4000, 5000, 6000, 7000, 8000] K, 
and mass densities of [0.3, 0.4, 0.6, 0.8, 1.0, 1.2, 1.4] \si{g/cm^3}. 
The conditions for high density include temperatures of [1000, 1500, 2000] \si{K}, 
and mass densities of [1.8, 2.2, 2.6] \si{g/cm^3}. 
In total, this covers 51 thermodynamic state points. Each simulation lasts 3 ps. 
The \textit{NVT} ensemble with Nos\'e-Hoover thermostat is used for the AIMD simulations.
The time step of 1 \si{fs} is used throughout the whole study. 

The total number of atoms in the AIMD simulation is 64 for the low density systems 
and 108 for the high density systems, where a $\Gamma$-point 
sampling is sufficient for the Brillouin zone integration. 
To properly account for the weak interaction 
in the vapor phase, we use the GGA functional PBE \cite{PBE1996} with the  
D3 dispersion correction, denoted PBE-D3, as the exchange-correlation 
functional.

For the structural comparison, 
we selected a few conditions to run 10 \si{ps} simulations with 108 atoms.
We extract the radial distribution function (RDF) from the AIMD simulations after
equilibration.
After collecting representative configurations, 
we perform self-consistent field (SCF) calculations to obtain 
accurate energies, forces, and virials for each functional.

\subsection{Training and validation of the deep potential}
\label{sec:nnpot}
Following the Behler-Parrinello approach\cite{Behler2007, Behler2011}, the DP used here is 
a machine-learning many-body interatomic potential
trained on the \textit{ab initio} data, providing both efficiency and accuracy for
large-scale molecular dynamics simulations. 
The DP model is trained using the DeepPot-SE (DP Smooth Edition)
model in DeePMD-kit via DPGEN \cite{DeepMD2018,DPGEN2020}, ``se\_e2\_a", 
where both angular and radial components of the atomic configurations are used to 
represent the atomic environment. 
Here ``e2"  stands for the two-atom embedding descriptor, which is 
quite suitable for rather dilute systems near the critical point. 
The potential smoothness cutoff radius is set to 6.0 \si{\angstrom}, 
and is sufficient to accurately describe the structures of 
liquid and vapor phases.
See Supplemental Material Section S1 at \cite{SuppMaterial} for 
detailed settings of the DPGEN iterations.

To assess the accuracy of the DP, we evaluate the
model on AIMD data. The test dataset is composed of frames that are 
not included in the training dataset.
The root mean square errors (RMSEs) of the energies, forces, and virials
are \SI{9.7}{meV/atom}, \SI{0.0614}{eV/\angstrom}, and \SI{19.0}{meV/atom}, respectively.
The errors are quite small, indicating that the DP is accurate enough to be used for
further simulations. With this, we perform deep potential molecular dynamics (DPMD) simulations using the 
LAMMPS package \cite{LAMMPS}.
The RDFs from the DPMD simulations are in excellent agreement with the AIMD results,  
indicating that the DP is able to accurately reproduce the structural properties of liquid and vapor phases.
In addition, the pressures from the DPMD simulations are also in good agreement with the AIMD results;
see Supplemental Material Section S2 at \cite{SuppMaterial} for details on these comparisons.
\subsection{XC functional calibration}
\label{sub:xc_calibration}

Ideally, the exact XC functional would deliver both the liquid and vapor densities all the way to the critical point and beyond.
However, only approximations to XC functionals are available in practice.
Thus, a judicious choice of XC functional is important for accurately describing liquid-vapor 
coexistence. 
Following the universal scaling law (Eq. \ref{eq:scaling}) and the law of rectilinear diameters (Eq. \ref{eq:rectilinear}),
the property that must be described most accurately is the liquid density at a given temperature under ambient pressure,
because of the large disparity between liquid density and vapor density ($\rho_l \gg \rho_v$ when $T \ll T_c$). In addition,
the interaction in the vapor phase is weak, and the vapor density is very low, which makes it less sensitive to the choice of XC functional. 
In contrast, the many-body interaction in the liquid phase is stronger and more intricate, 
which is harder to capture with empirical or classical interatomic potentials. This is where the neural-network representation shines,
as it can capture the complex many-body interactions in the liquid phase. 

\begin{figure}[htbp!]
\centering
\includegraphics[width=0.7\textwidth]{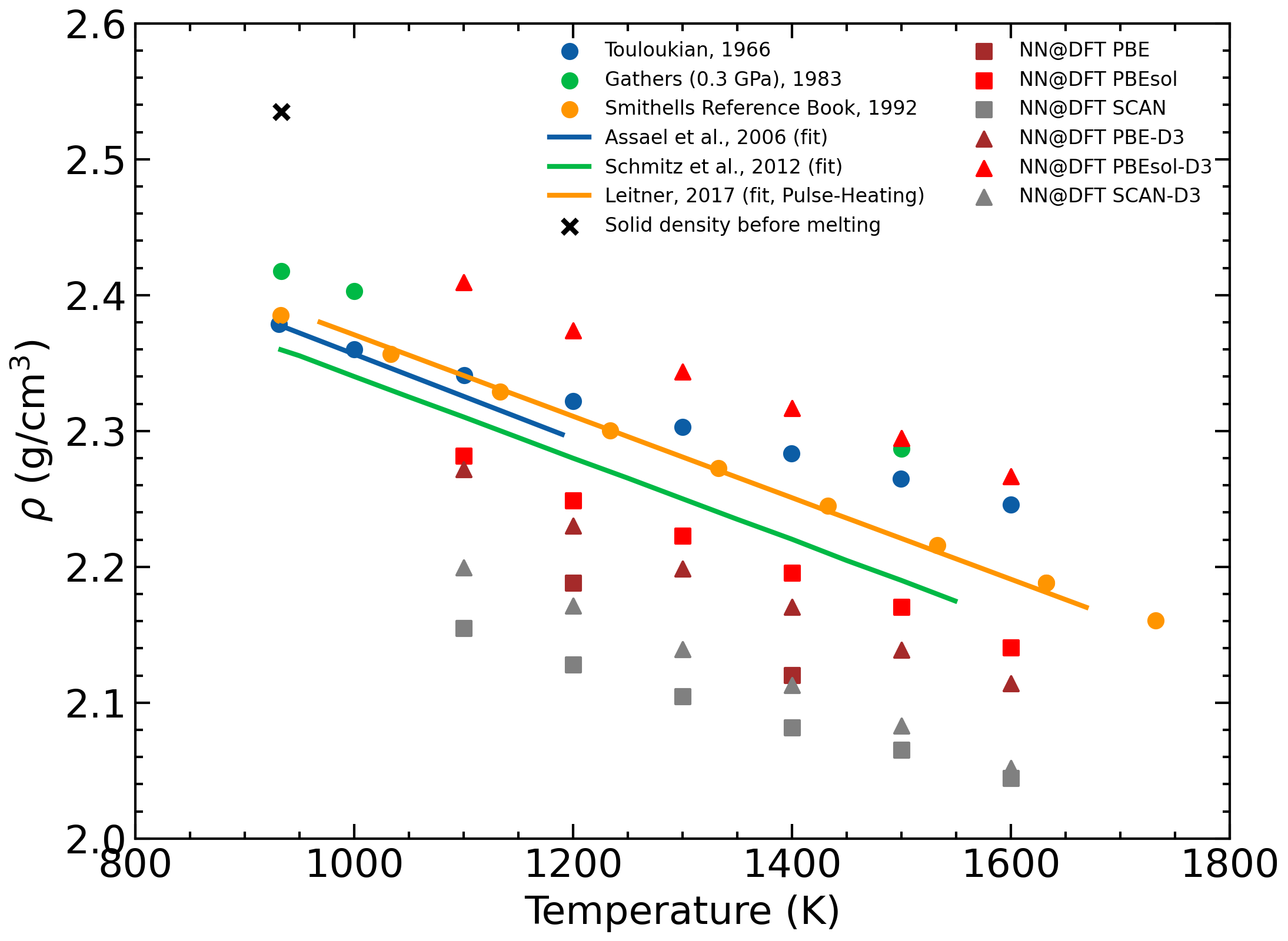}
\caption{Bulk liquid density of Al at different temperatures from experiments and theoretical calculations at ambient pressure. 
The experimental data are from Touloukian \cite{Touloukian1966} (blue dot), Gathers \cite{Gathers1983} (green dot), Smithells \cite{Smithells2013} (orange dot), 
Assael \cite{Assael2006} (blue line), Schmitz \cite{Schmitz2012} (green line), and  pulse-heating data of Leitner\cite{Leitner2017} (yellow line).
The solid density before melting is also included as a reference (black cross) \cite{Touloukian1966}. 
The theoretical calculations are obtained using neural networks trained on \textit{ab initio} data using different XC functionals, 
PBE \cite{PBE1996} (brown square), PBEsol\cite{PBEsol2008} (red square), and SCAN\cite{Sun2015} (grey square) and their dispersion corrected counterparts (colored triangles). 
The B3LYP hybrid functional data are omitted for 
significantly underestimating the density. Overall, PBEsol provides the best agreement with 
the experimental data.
}
\label{fig:density}
\end{figure}

Using \textit{NPT} molecular dynamics (MD) simulations, 
we compute the liquid density at 1 bar for temperatures ranging from 
1100 \si{K} to 1600 \si{K} using various neural-network potentials distinguished by XC functional.
The experimental densities are displayed in Fig. \ref{fig:density} 
for comparison \cite{Touloukian1966, Gathers1983,Smithells2013, Assael2006,Schmitz2012, Leitner2017}.
Because it significantly underestimates the density, the B3LYP hybrid functional \cite{Becke1993,Lee1988} is 
omitted from the figure despite significant effort devoted to achieving convergence. 
The meta-GGA functional SCAN \cite{Sun2015} and the GGA functional PBE\cite{PBE1996} still underestimate the density, 
although the discrepancy is less severe than for B3LYP. Inclusion of the van der Waals interaction via DFT-D3 \cite{Grimme2010,Grimme2011} 
improves the agreement by bringing the predicted densities closer to the experimental values, but the density is still underestimated. 
In contrast, the PBEsol functional was explicitly designed to restore the density-gradient expansion for exchange, 
making it accurate for slowly varying densities that dominate both bulk liquids and interfacial regions. 
The jellium surface energy benchmark---directly relevant to liquid-vapor interfaces---shows that PBEsol reduces the surface exchange energy
 error from 11\% (PBE) to 2.7\% \cite{PBEsol2008}, demonstrating superior performance in the low-density tail region. 
The predicted density lies closer to the experimental liquid-density data and remains within the experimental error bars.
Including the dispersion correction to PBEsol, denoted as PBEsol-D3, leads to an overestimation of the density.
The PBEsol liquid density has the correct curvature with temperature, indicating that the thermal expansion coefficient is also well captured. 
For the vapor phase, despite not being explicitly calibrated, the much larger thermal expansion coefficient is also captured well by PBEsol, as evidenced by the Clausius-Clapeyron relation (see Fig. \ref{fig:clausius}).
Overall, PBEsol provides a reliable description of both liquid and vapor phase behavior in aluminum. The following calculations are therefore based on PBEsol.

\subsection{Temperature quench molecular dynamics }
\label{sub:tqmd}
In conventional direct coexistence MD simulations, previously equilibrated bulk
liquid and bulk vapor configurations are placed in a simulation box, typically in the form of a liquid slab surrounded by
vapor. The system is then allowed to evolve under \textit{NVT} conditions until equilibrium is reached
through diffusive mass transport. However, this approach is 
less efficient because long simulation times are needed to equilibrate the two-phase system. 

Instead, we adopt the approach of temperature quench molecular dynamics 
(TQMD) \cite{Gelb2002}. In TQMD, we first equilibrate the system at a 
high temperature, then rapidly lower the temperature---either instantly or 
over a short time---by rescaling atomic velocities or adjusting the 
thermostat, and finally let the system evolve at the new temperature to 
observe structural changes. The sudden temperature quench
places the system in a thermodynamically and mechanically unstable state, namely 
a supercooled state.

The phases are identified and characterized by partitioning the system 
into small blocks and examining the densities and/or compositions within 
each block. This approach allows phase coexistence data to be obtained 
from a locally equilibrated system, eliminating the need to run the 
simulation until full global equilibrium is achieved, thereby greatly 
reducing computational cost. To identify the liquid and vapor phases and their densities, 
we develop a novel method based on Gaussian mixture model (GMM) fitting.
See Supplemental Material Section S1 at \cite{SuppMaterial} for 
details of the method.

\subsection{Critical point identification}
\label{sub:critical_point}
We use two complementary approaches to identify the critical point. From the EOS simulations, we find the critical point by fitting the
pressure-volume-temperature (\textit{PVT}) data to a model EOS.
The widely used simple third-order in density EOS model leads to large fitting errors. 
Instead, we find that a fourth-order EOS in density accommodates the data better. 
The EOS model is given by
\begin{equation}
    P(T,\rho) = (a_0 + a_1 T) \rho + (b_0 + b_1 T) \rho^2 + (c_0 + c_1 T) \rho^3 + (d_0 + d_1 T) \rho^4
    \label{eq:fourth_order}
\end{equation}
where $a_0, a_1, b_0, b_1, c_0, c_1, d_0, d_1$ are fitting parameters.
Then the critical point can be found from the diverging isothermal compressibility
\begin{equation}
\left. \frac{\partial P}{\partial \rho} \right|_{T_c, \rho_c} = 0, \quad
\left. \frac{\partial^2 P}{\partial \rho^2} \right|_{T_c, \rho_c} = 0
\label{eq:critical}
\end{equation}
where $P$ is the pressure and $\rho$ is the density.
The critical temperature $T_c$ and critical density $\rho_c$ can be obtained by 
solving Eq. \ref{eq:critical} simultaneously for $T_c$ and $\rho_c$. 

For the liquid-vapor coexistence approach from TQMD, 
the liquid-vapor critical point can be obtained by 
fitting the coexistence densities of the liquid and vapor phases 
to the universal scaling law of coexistence densities and the law of rectilinear diameters \cite{RowlinsonWidom, Statt2020}. 
The universal scaling law of coexistence densities is given by:
\begin{equation}
\rho_l - \rho_v = A(T_c - T)^{\beta},
\label{eq:scaling}
\end{equation}
where $\rho_l$ and $\rho_v$ are the densities of the liquid and vapor phases, respectively, 
$T_c$ is the critical temperature, $A$ is a fitting parameter, and $\beta$ is the critical exponent, 
which is typically around 0.326 for three-dimensional systems\cite{ZinnJustin2001}.
The law of rectilinear diameters is given by:
\begin{equation}
\frac{\rho_l + \rho_v}{2} = \rho_c + B(T_c - T),
\label{eq:rectilinear}
\end{equation}
where $\rho_c$ is the critical density, and $B$ is another fitting parameter.

By fitting the coexistence densities to these two equations, 
we can extract the critical temperature and density. 
To extract the infinite system size critical point, we perform
finite-size scaling analysis. The finite-size scaling relation is given by:
\begin{eqnarray}
  T_c(L) - T_c(\infty) \sim L^{-(1+\theta)/\nu} \\
  \rho_c(L) - \rho_c(\infty) \sim L^{-( 1 - \alpha)/ \nu } 
\end{eqnarray}
where $\theta$, $\nu$, and $\alpha$ are the correction-to-scaling exponent, the critical exponent of the correlation length, and
the exponent associated with the heat capacity divergence, respectively. 
For the 3D Ising universality class, $\theta \approx 0.54$, $\nu \approx 0.63$, and $\alpha \approx 0.11$\cite{Sengers1986,Liu1989}.

\section{Results and Discussion}
\label{sec:results}
To give a brief overview, the critical parameters (temperature, density, pressure, and compressibility factor) for aluminum have been summarized in 
chronological order in Table~\ref{tab:critical_summary}. 
Young and Alder showed that, using a modified van der Waals model, the CP
of aluminum and other metals can be semi-quantitatively predicted, 
with $T_c = 7151$ \si{\si{K}}, $\rho_c = 0.69$ \si{g/cm^3}, and  $P_c = 5.458$ \si{kbar} \cite{Young1971}.
Likalter incorporated the plasma-like features of metallic elements 
close to the CP using a percolation model and predicted  $T_c = 8860$ \si{K}, 
$\rho_c = 0.28$ \si{g/cm^3}, and  $P_c = 4.68$ \si{kbar} \cite{Likalter1996}.
Bhatt et al. used Gibbs ensemble Monte Carlo simulations with
embedded-atom method (EAM) potentials to estimate the critical point 
\cite{Bhatt2006}. 
However, these methods are limited by the accuracy of the interatomic
potentials used or the model employed. Desjarlais instead directly 
searches for the diverging isothermal compressibility characteristic
using long and demanding \textit{ab initio} molecular dynamics (AIMD) simulations\cite{Desjarlais2009}, and the obtained critical parameters were used 
by Lomonosov to construct the multiphase EOS\cite{Lomonosov2007}. 
Similarly, Faussiurier et al. used hard-sphere \textit{ab initio} molecular dynamics and a fit to
a cubic EOS to estimate the critical point\cite{Faussurier2009} with 
$T_c = 7953$ \si{K}, $\rho_c = 0.44$ \si{g/cm^3}, $P_c = 0.35$ \si{kbar}. 
However, the system sizes and timescales accessible to \textit{ab initio} simulations render an 
accurate determination of the critical point extremely challenging. Next, we present
our results from both the EOS and TQMD approaches, which converge on a consistent critical point with much smaller uncertainty than previous estimates.

\begin{table}[htb!]
\caption{Summary of reported critical parameters for aluminum. 
Blank entries denote values not provided in the cited reference.
$Z_c$ is the critical compressibility factor.
}

\label{tab:critical_summary}
\begin{ruledtabular}
  \begin{tabular}{lcccccl}
    Reference & Year & $\rho_c$ (g/cm$^{3}$) & $T_c$ (K) & $P_c$ (kbar) & $Z_c$ & Methodology \\
    \hline
    Young \cite{Young1971} & 1971 & 0.69  & 7151 & 5.458 & 0.36 & EOS \\
    Likalter \cite{Likalter1996} & 1996 & 0.28  & 8860 & 4.68  & 0.61 & EOS \\
    Hess \cite{Hess1998} & 1998 & 0.43  & 8944 & 4.726 & 0.40 & Vapor pressure curves \\
    Singh \cite{Singh2006} & 2006 & 0.785 & 8472 & 5.094 & 0.25 & Coexistence (Monte Carlo) \\
    Ray \cite{Ray2006} & 2006 & 0.32  & 5700 & 1.87  & 0.33 & EOS \\
    Lomonosov \cite{Lomonosov2007} & 2007 & 0.703 & 6250 & 1.97  & 0.15 & EOS \\
    Gordeev \cite{Gordeev2008} & 2008 & 0.66  & 7917 & ---    & ---  & EOS \\
    Povarnitsyn \cite{Povarnitsyn2008} & 2008 & 0.698 & 6595 & ---    & ---  & EOS \\
    Faussurier \cite{Faussurier2009} & 2009 & 0.44  & 7963 & 3.5   & 0.32 & EOS \\
    Leitner \cite{Leitner2017} & 2017 & 0.55  & 4500 & ---    & ---  & EOS \\
    This work & 2026 & 0.637 & 6539 & 1.61   & 0.125  & EOS \\
    This work & 2026 & 0.586 & 6576  & 1.67  & 0.14  & Coexistence (TQMD) \\
  \end{tabular}
\end{ruledtabular}
\end{table}

\subsection{Critical parameters from spinodal lines}
\label{sub:spinodal}
Close to the critical point, the system exhibits large density fluctuations. We have noticed that the pressure fluctuations
is significantly amplified for system sizes smaller than 1000 atoms, making it difficult to obtain a smooth EOS curve.
Such large simulations become infeasible for AIMD, but are easily accessible with the DP.
Therefore, we perform \textit{NVT} MD simulations with 4096 atoms to obtain the EOS isotherms for temperatures ranging from
5000 K to 8000 K. With these data, as noted above, the previously used simple third-order EOS model in density 
leads to large fitting errors, and we find that a fourth-order EOS model in density accommodates the data better.
See Supplemental Material at \cite{SuppMaterial} for fitting coefficients.
\begin{figure}[thbp!]
\centering
\includegraphics[width=0.8\textwidth]{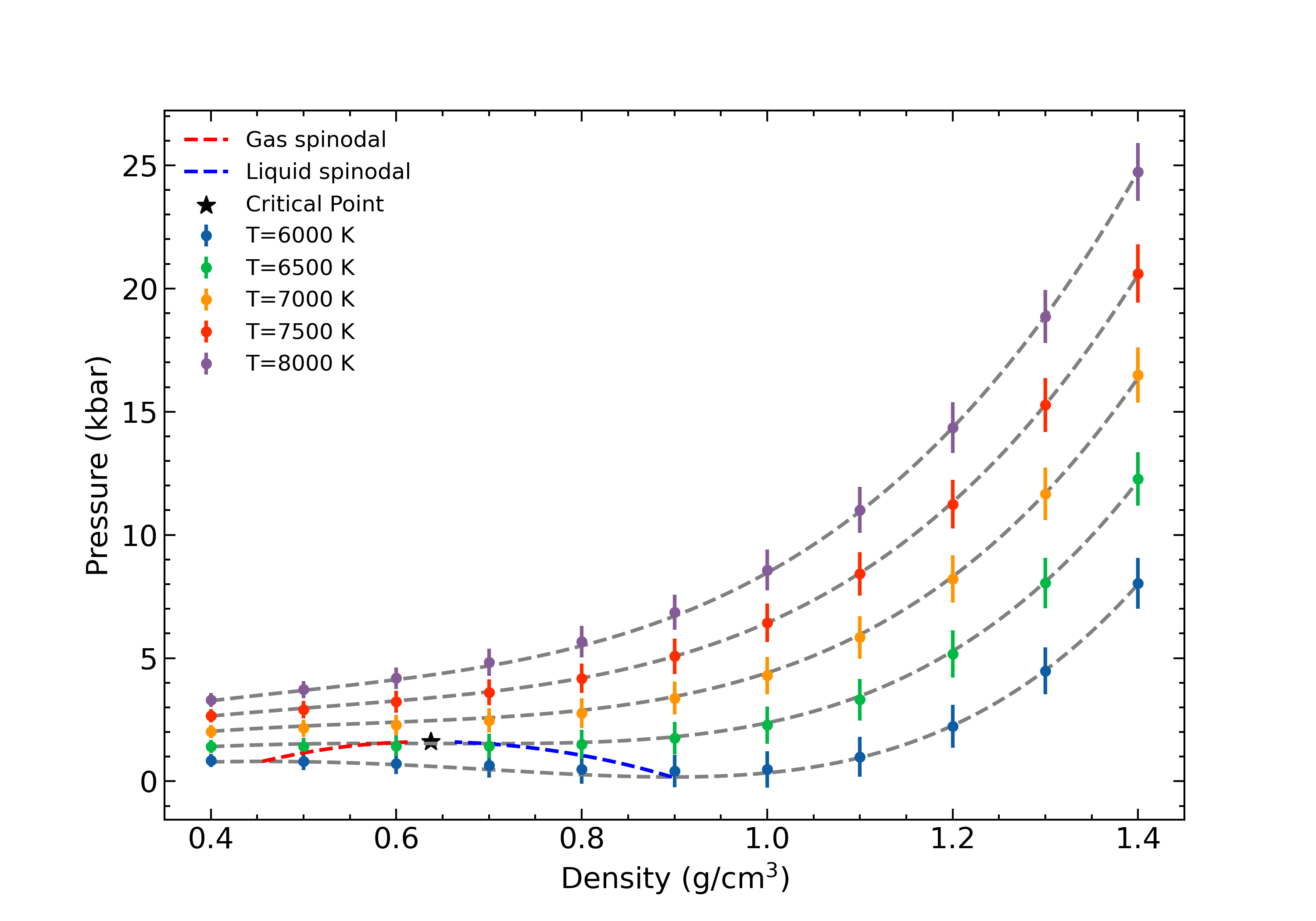}
\caption{Equation of state isotherms for Al using the DP. The standard deviation
has been significantly reduced (see the error bar on each data point).
A fit to the fourth-order EOS in density 
(see text) yields the liquid (blue dashed line) and vapor spinodal (red dashed line) lines, which meet at the critical point (marked by a black star). This CP is identified at $T_c = 6539$ \si{K},
$\rho_c = 0.637$ \si{g/cm^3}, and $P_c = 1.6$ \si{kbar}.
}
\label{fig:eos}
\end{figure}

The critical pressure $P_c$ is then obtained by substituting $T_c$ and $\rho_c$ back into Eq. \ref{eq:fourth_order}.
Taking into account the uncertainty in the EOS fitting parameters, we properly propagate the errors to obtain the uncertainty in the critical parameters.
We build both liquid and vapor spinodal lines forming a dome in the $P-\rho$ plane (see Fig. \ref{fig:eos}), where 
the critical point is located at the apex of the dome.
We obtain $T_c = 6539 \pm 151$ \si{K}, $\rho_c = 0.637 \pm 0.006$ \si{g/cm^3} , and $P_c = 1.61 \pm 0.25$ \si{kbar} (see Fig. \ref{fig:eos}).
Previous estimates by Desjarlais \cite{Desjarlais2009} using AIMD
 gave $T_c \approx 6000$ \si{K} with 10\% uncertainty, $\rho_c = 0.63$ \si{g/cm^3}, and $P_c \approx 1.8$ \si{kbar}.
The critical density and pressure are in good agreement, but our $T_c$ is about  540 K higher. This difference is 
likely due to the choice of XC functional, probably PBE although the functional was not explicitly stated.
It is worth noting, however, that 
our $T_c$ lies within the uncertainty range of Desjarlais' estimate, while our uncertainty is much smaller, at about 150 \si{K}.
Our uncertainty of about 2.2\% is much smaller than the previous estimate of 10\%.
The compressibility factor at the critical point is calculated from $Z_c = P_c / (\rho_c R_{sp} T_c)$,
with $R_{sp}$ being the specific gas constant.
It is significantly smaller than the ideal 
gas value of 1.0, indicating strong interactions in the system at the critical point.
Our predicted compressibility factor around 0.13 is smaller than the previous estimates and is close to 
the value of 0.15 given by Lomonosov,
which is largely attributable to the critical pressure being smaller than previous estimates.
It is worth noting that a shallow van der Waals loop  forms in the EOS isotherms below $T_c$.






\subsection{Critical parameters from liquid-vapor coexistence}
\label{sub:coex}
In order to obtain a liquid-vapor coexistence curve, typically a few thousand atoms 
are required to simulate the two-phase system on the timescale of sub-nanoseconds.
With the DP, we can directly simulate such large systems.
 We employ the temperature quench molecular dynamics (TQMD) to obtain the liquid-vapor coexistence curve (see Section \ref{sub:tqmd}).
We choose the box dimensions such that $L_x = L_y$ and $L_z = 5 L_x$. Depending on the number of atoms placed in the box, 
the box dimensions scale proportionally. 
An initial density of 0.6\unit{g/cm^3}, not far from the expected critical density, is used to equilibrate the system 
at a much higher temperature (12000 \si{K}) in the \textit{NVT} ensemble. 
Then the system is suddenly quenched to a target temperature, 
e.g. 4500 \si{K}, and then equilibrated for 500 \si{ps}. During this stage, the liquid-vapor interface gradually forms
and stabilizes (see the gif file in the Supplementary Data \cite{Zenodo}). 
Finally, a production run of 50 \si{ps} is performed to collect the data for analysis.

In the TQMD approach, we develop a microstructural analysis method 
based on Gaussian mixture model (GMM) fitting
to identify the liquid and vapor densities.
Each atom in the system is assigned to either the liquid or vapor phase based on its local atomic environment
(see Fig. S4 in Supplemental Material at \cite{SuppMaterial}).

The total volume occupied by the liquid (vapor) phase in the Voronoi tessellation is then calculated,
and the corresponding density is obtained straightforwardly.
After collecting the density-evolution data, we perform a statistical analysis to obtain the average density and its 
 uncertainty for both phases at each temperature. 
See Supplemental Material Section S3 and figures at \cite{SuppMaterial} for details of the method.

\begin{figure}
\centering
\includegraphics[width=0.8\textwidth]{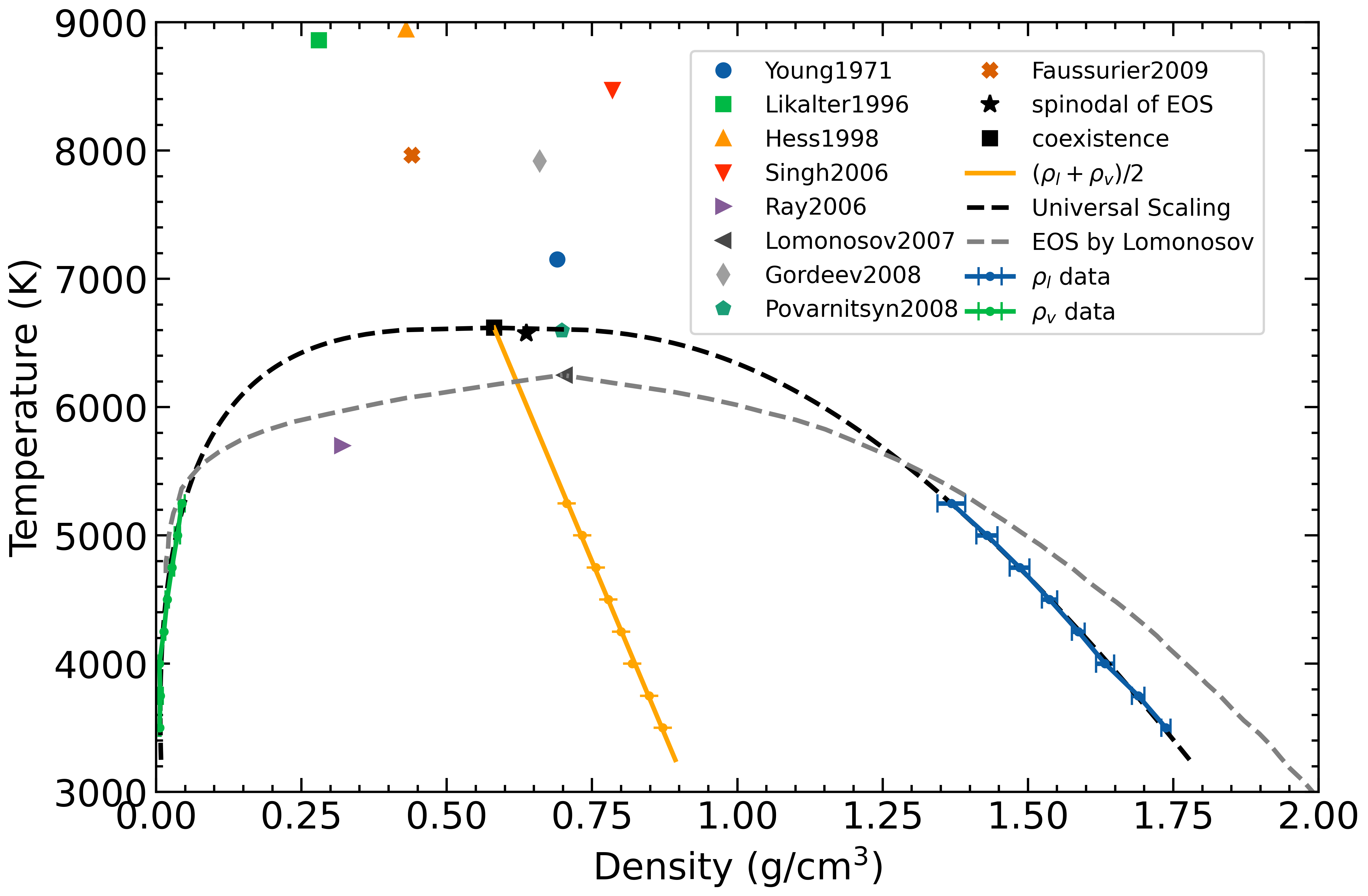}
\caption{Liquid-vapor coexistence curve for Al obtained from DP temperature 
quench molecular dynamics with PBEsol. 
The phase-identification analysis yields the densities and 
their uncertainties for the bulk liquid (blue) and bulk vapor (green) 
at each temperature along the coexistence curve. The total number of atoms is 4000.
Fitting the coexistence densities to the universal scaling law and the 
law of rectilinear diameters identifies 
the critical point (marked by a black square) at
$T_c = 6618 \pm 23$ \si{K} and $\rho_c = 0.5814 \pm 0.004$ \si{g/cm^3}.
The critical point from spinodal analysis is also shown (black star).
Early predictions of the critical point are shown as scattered symbols,
 e.g. Young \cite{Young1971} (circle), Likalter
 \cite{Likalter1996} (square), Hess\cite{Hess1998} (upper triangle),
 Singh\cite{Singh2006}(down triangle), Ray\cite{Ray2006} (right triangle), 
 Lomonosov\cite{Lomonosov2007} (left triangle), 
 Gordeev\cite{Gordeev2008} (diamond), Povarnitsyn
 \cite{Povarnitsyn2008} (pentagon), and Faussurier\cite{Faussurier2009}(cross).
}
\label{fig:coex}
\end{figure}

As mentioned earlier, the location of the liquid-vapor critical point is poorly constrained 
(as illustrated by the wide spread of points in Fig. \ref{fig:coex}).
To identify the critical point, we adopt the universal scaling law and the rectilinear diameter law \cite{RowlinsonWidom,Statt2020}. 
Metallic fluids, despite long-range Coulomb correlations, are widely accepted to follow Ising universality due to screening,
and we find that the 3D Ising critical exponents fit our data well 
(see Fig. S6 in Supplemental Material at \cite{SuppMaterial}).
 With these, we obtain the liquid-vapor coexistence curve for aluminum, as shown in Fig. \ref{fig:coex}.

Because the critical temperature depends on system size,
we perform the same analysis for systems containing $2000$, $3000$, and $4000$ atoms.
These correspond to $L_z$ values of $155.136$, $177.586$, and $195.45$ \AA, respectively.
Finite-size scaling theory \cite{Wilding1995} predicts how the critical temperature and critical density deviate from their infinite-size  
limits. 
The infinite-size  
limit is $T_c(\infty) = 6576$ \si{K}. 
However, for the critical density, 
we find it fluctuates around 0.586 \si{g/cm^3}
without a consistent scaling behavior. Substituting this value into the EOS, we obtain  the critical pressure $P_c = 1.67$ \si{kbar}
and compressibility factor $Z_c = 0.14$.

In Fig. \ref{fig:coex}, we compare our liquid-vapor coexistence curve  with previous predictions by 
 Lomonosov\cite{Lomonosov2007}.
 The vapor density data from our simulation  
compared to Lomonosov's are in good agreement. 
At the same time, the liquid density data from our simulation is evidently lower. 
Their curve agrees well with the experimental points Gathers \cite{Gathers1983}, which 
as we have shown in Fig. \ref{fig:density}, is rather an overestimation over a range of 
experimental consensus. Nevertheless, their extrapolation to the critical point is less transparent
and depends strongly on the EOS model employed, whereas ours relies on the universal scaling law.


Compared to early predictions of the critical point, our predicted critical temperature is close to the 
value of Povarnitsyn \cite{Povarnitsyn2008}, where $T_c = 6595 \pm 30 $ \si{K}. 
However, the predicted critical density 
is lower than their value of $0.6978$ \si{g/cm^3}. Based on statistical analysis, Morel et al. suggested
$T_c = 6700 \pm 800 $ \si{K} and $\rho_c = 0.566$ \si{g/cm^3} \cite{Morel2009}, which is 
in close agreement with our values.

From the EOS isotherms in Fig. \ref{fig:eos},
$T_c = 6539 \pm 151 $ \si{K}, $\rho_c = 0.637 \pm 0.006$ \si{g/cm^3}, $P_c = 1.61 \pm 0.25 $ \si{kbar}, and $Z_c = 0.125$ are obtained, 
which are generally consistent with the coexistence results. 
This discrepancy in density may arise from the different methods employed. 
Combining the diverging isothermal compressibility from EOS and the coexistence simulation, 
we find that the resultant 
critical temperature can be bracketed between 6539 \si{K} and 6576 \si{K}. 
Since the critical density is not well behaved in the finite-size scaling analysis,
we thus take the EOS result of 0.637 \si{g/cm^3} as the final density estimate. 
The critical pressure can be estimated as $P_c = 1.6$ \si{kbar}.


\subsection{Vapor pressure}
\label{sub:vapor_pressure}
\begin{figure}
\centering
\includegraphics[width=0.8\textwidth]{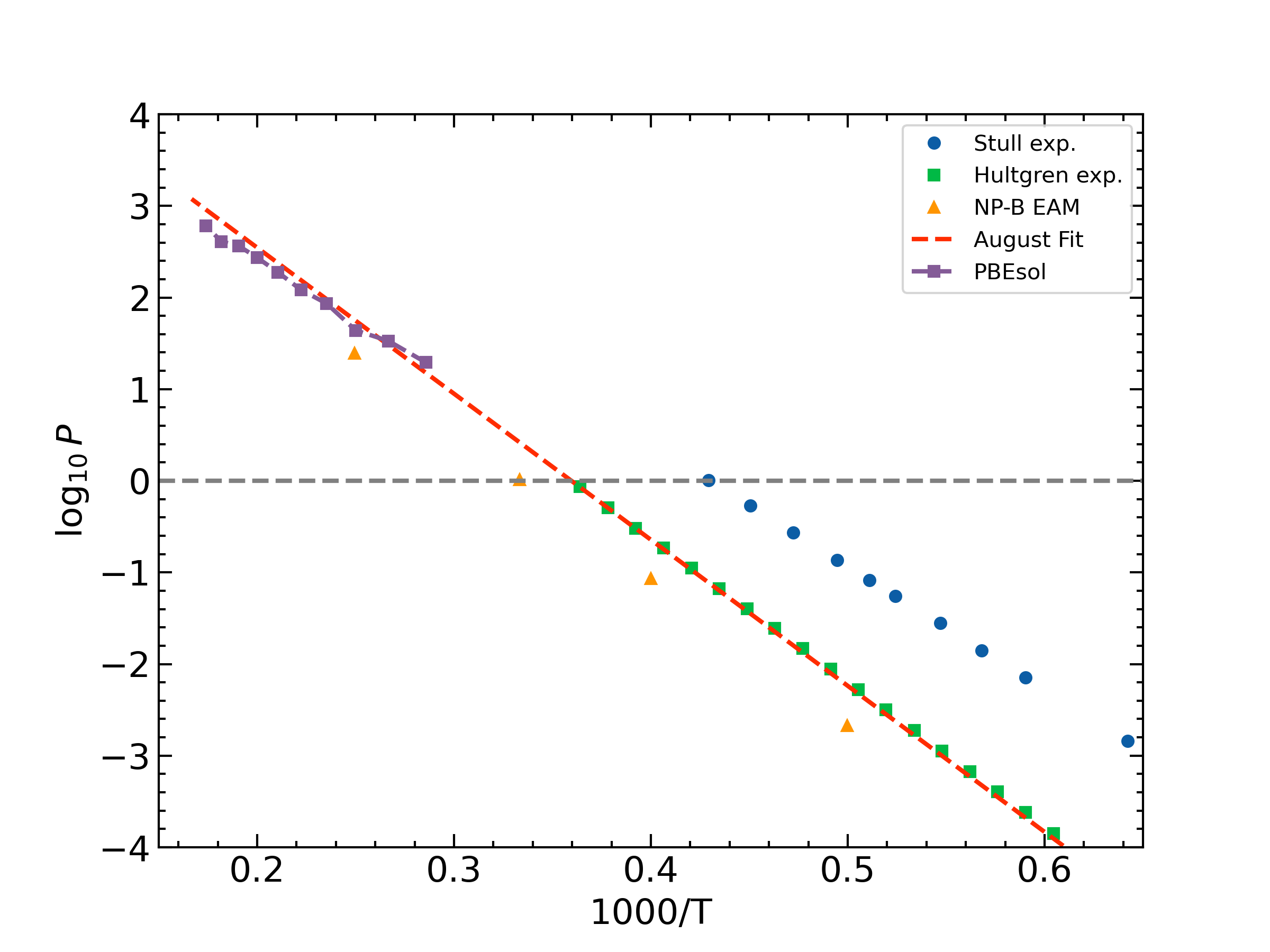}
\caption{
Vapor pressure-temperature relation for Al obtained from liquid-vapor coexistence simulations. 
Early experimental data are from Stull \cite{Stull1947} (blue dot), which deviate substantially from 
the later, more accurate data from Hultgren \cite{Hultgren1973} (green square). 
Using Hultgren data, the August equation (green line) is fitted to extrapolate beyond 
the boiling point at 1 atm. Theoretical calculations with PBEsol (purple square) align with the 
August line well. The EAM potential NP-B (yellow triangle) from Bhatt \cite{Bhatt2006} significantly
underestimates the vapor pressure.
}
\label{fig:clausius}
\end{figure}

The Clausius-Clapeyron equation describes the relationship between the pressure 
and temperature of a phase transition. From the liquid-vapor coexistence simulations, 
we can directly extract the vapor pressure at each temperature.
From the coexistence-curve simulations, we obtain the vapor pressure-temperature relationship,
which is shown in Figure \ref{fig:clausius}. 

The vapor pressure-temperature curve in Fig. \ref{fig:clausius},
shows good agreement with the experimental data from Hultgren \cite{Hultgren1973}. 
Early experimental data from Stull \cite{Stull1947} deviate substantially from 
the later, more accurate data from Hultgren \cite{Hultgren1973}.
By fitting the August equation to the Hultgren data, we find that our vapor pressure 
data from the TQMD simulations with PBEsol agree well with this fit. This 
agreement is not imposed a priori and thus provides a strong validation of our simulation approach.
The boiling point estimated from our simulation is very close to the experimental value, 
which again validates the accuracy of our approach.
This result greatly improves on the previous 
prediction from an embedded-atom method (EAM) potential, NP-B \cite{Bhatt2006}.


\section{Summary}
\label{sec:summary}

In summary, we have resolved the long-standing uncertainty in the location of the liquid--vapor critical point of aluminum by combining deep 
potential molecular dynamics with large-scale simulations at \emph{ab initio} accuracy. Through a systematic assessment of 
exchange--correlation functionals and two complementary approaches---spinodal analysis of the EOS and direct 
coexistence simulations---we converge on a critical temperature of $6531$--$6576$~K, a critical density of $0.637$~g/cm$^{3}$, and 
a critical pressure of $1.6$~kbar, with temperature uncertainties below 50~K.
Finite-size scaling analysis further supports the robustness of our critical point estimates, 
and the vapor pressure-temperature relationship aligns well with experimental data, 
confirming the reliability of our methodology.
This level of precision represents a decisive 
advance over previous estimates and provides a benchmark for both theory and experiment.

Our study relies on critical exponents established for model systems. Despite the power of deep potentials, direct simulations close to the critical point remain challenging 
because of the intrinsically large fluctuations. Further increases in system size, enabled by graphics processing units, 
may be necessary to reduce the uncertainty further and eventually resolve the critical exponents directly. 
Beyond aluminum, our work establishes a transferable framework for predicting critical phenomena in metals and complex materials 
under extreme conditions. The combination of high-fidelity machine-learning potentials, robust coexistence analysis, and 
finite-size scaling offers a general strategy for mapping phase diagrams with unprecedented accuracy. These results open the way 
to quantitative modeling of laser ablation, shock compression, and planetary matter, and underscore the broader role of 
data-driven atomistic simulations in advancing high-energy-density physics.

\subsection*{Acknowledgments} 
This work received no external funding. We
 thank S.B. Trickey for detailing the history of this problem.
We would like to express our gratitude to Ronald E. Cohen for 
helpful discussions. 
We thank anonymous reviewers for their constructive comments that have significantly improved the quality of this work.

\subsection*{Data Availability}
The training set, the potential models, the input scripts, the figure files, as well as 
the resulting DP model can be found at Zenodo
\cite{Zenodo}. 
Because of their large size, the MD trajectories are stored on a local server and are available from the 
corresponding author upon reasonable request.
Additional data are available from the corresponding author upon reasonable request.



%

\end{document}